\documentclass[twocolumn,showpacs,preprintnumbers,amsmath,amssymb]{revtex4}
\usepackage{graphicx}% Include figure files
\usepackage{dcolumn}% Align table columns on decimal point
\usepackage{bm}% bold math
\usepackage{colordvi}

\begin{document}

\preprint{SFB393}

\title{Lyapunov instabilities of Lennard-Jones fluids}% Force line breaks with \\

\author{Hong-liu Yang}
 \email{hongliu.yang@physik.tu-chemnitz.de}
\author{G\"unter Radons}%
 \email{radons@physik.tu-chemnitz.de}
\affiliation{%
Institute of Physics\\
Chemnitz University of Technology, D-09107 Chemnitz, Germany 
}%

%\affiliation{

\date{\today}
\begin{abstract}
Recent work on many particle system reveals the existence
of regular collective perturbations corresponding to the smallest
positive Lyapunov exponents (LEs), called hydrodynamic Lyapunov
modes. Until now, however, these
modes are only found for hard core systems. Here we report new
results on Lyapunov spectra and Lyapunov vectors (LVs) for 
Lennard-Jones fluids. 
By considering the Fourier transform of the coordinate fluctuation density $u^{(\alpha)}(x,t)$,
it is found that the LVs with $\lambda \approx 0$ are highly
dominated by a few components with low wave-numbers. 
 These numerical results provide strong evidence
that hydrodynamic Lyapunov modes do exist in soft-potential systems,
although the collective Lyapunov modes are more vague than in hard-core systems.
In studying the density and temperature dependence of these 
modes, it is found that, when the value of Lyapunov exponent $\lambda^{(\alpha)}$ is plotted as function of the dominant wave number $k_{max}$ of the corresponding LV, 
all data from simulations with different densities and temperatures collapse onto a single curve. This shows that the dispersion relation $\lambda^{(\alpha)}$ vs. $k_{max}$
for hydrodynamical Lyapunov modes appears to be universal irrespective of the particle density and temperature of the system. 
 Despite the wave-like character of the LVs, no step-like structure exists in the Lyapunov spectrum of the systems studied here, in contrast to the hard-core case. 
Further numerical simulations show that the finite-time LEs fluctuate strongly.  
We have also investigated localization features of LVs and propose a new length scale to characterize the Hamiltonian spatio-temporal 
chaotic states.
\end{abstract}



\maketitle

\section{\label{sec:level1}Introduction}One of the most successful theories in modern science is statistical
mechanics, which allows one to understand the macroscopic (thermodynamic) properties of matter from a
statistical analysis of the microscopic (mechanical) behavior of the constituent particles. In spite of this, using certain probabilistic
assumptions such as Boltzmann's {\it Stosszahlansatz} renders the lack of a firm foundation of this theory, 
especially for non-equilibrium statistical mechanics. Fortunately, 
the concept of chaotic dynamics developed in the 20th century \cite{EckmannRuelle} is a good candidate for complying with these difficulties. Instead of 
the probabilistic assumptions, the dynamical
instability of trajectories can provide the necessary fast loss of time correlations, ergodicity, mixing and other
dynamical randomness \cite{krylov}. It is generally expected that dynamical instability is at the origin of macroscopic
transport phenomena and that one can find certain connection between them. In the past decade, some beautiful theories 
in this direction were already developed. Examples are the escape-rate formalism by Gaspard and Nicolis \cite{gaspard,dorfman} and the Gaussian 
thermostat method due to Nos\'e, Hoover, Evans, Morriss and others \cite{evans,hooverbk1,hooverbk2}, 
where the Lyapunov exponents were related to certain transport coefficients.

Very recently, molecular dynamics simulations on hard-core systems revealed the existence
of regular collective perturbations corresponding to the smallest
positive Lyapunov exponents (LEs), named hydrodynamic Lyapunov
modes \cite{posch-hirschl}. This opens a possible new way for the connection between Lyapunov vectors, a
quantity characterizing dynamical instability of trajectories, and
macroscopic transport properties. A lot of work \cite{PoschForster,posch2002,eckmann,france,wijin,Taniguchi,morriss,hoover} 
has been done to identify this phenomenon and to find
out its origin. It is commonly thought that the appearance of these modes is due to the conservation of certain
quantities in the systems studied \cite{posch2002,eckmann,france,wijin,Taniguchi}. 
A natural consequence of this expectation is that the appearance of such modes would not be an exclusive feature
of hard-core systems and should be generic to a large class of Hamiltonian systems. However, till now, these
modes were only identified in the computer simulations of hard-core
systems \cite{PoschForster,posch2002,hoover}.

In this work, we report on new results about a 1d system with Lennard-Jones interaction. Although the
identification of regular hydrodynamic Lyapunov modes by naked eye is difficult for soft potential systems \cite{hoover}, 
our new technique based on a spectral analysis of LVs shows strong evidence that hydrodynamic Lyapunov modes do exist
in this case. The influence of density and temperature changes is studied in detail. The dispersion relation for hydrodynamic Lyapunov modes,
the dominant wave-number as function of the corresponding LEs is found being universal for all densities and temperatures.

Furthermore we study the localization properties of LVs. Based on the extensive nature of LVs with $\lambda \approx 0$, we propose a new length
scale to characterize a spatio-temporal chaotic Hamiltonian system.  
It is expected that this new quantity will be useful for the task of distinguishing different
spatio-temporal chaotic states and characterizing transitions among them. This is an important open question 
in the study of spatial-temporal chaos \cite{mcross}.

\section{\label{sec:level2}Model}
In this study we use a 1D Lennard-Jones system with Hamiltonian
\begin{equation}
H=\sum_{i=1}^N mv_i^2/2+\sum_{i<j}V(x_{j}-x_{i})
\end{equation}
The interaction potential among particles is of the form:
\begin{equation}
V(r)=
\begin{cases}
4\epsilon\left[ (\sigma / r) ^{12}-(\sigma / r)^{6}\right ]-V_c  & \text{if $r \leq r_c$},\\
 0 & \text{otherwise}. 
\end{cases}
\end{equation}
with $V_c=4\epsilon\left[ (\sigma / r_c) ^{12}-(\sigma / r_c)^{6}\right ]$. The potential is truncated in order to
lower the computational burden. Other types of potential with smoothed force at the truncation point were also simulated to
check the influence on the results given below. No qualitative difference is found between them.

The system is integrated using the velocity form of the Verlet algorithm with
periodic boundary conditions \cite{kob}. In our simulations, we set $m=1$, $\sigma=1$, $\epsilon=1$ and $r_c=2.5$. All results
are given in reduced units, i.e., length in units of $\sigma$, energy in units of $\epsilon$ and time in units of
$(m\sigma^2/48\epsilon)^{1/2}$. The time step used in the molecular dynamics simulation is $h=0.008$. 
The standard Gram-Schmidt re-orthonormalization algorithm \cite{benettin,shimada} is used
to calculate the local dynamical instability of the systems studied. 
The time interval for periodic re-orthonormalization is $30h$ to $100h$.
Throughout this paper, the particle number typically is denoted by $N$, the length of the system by $L$ and the temperature by $T$.

\section{\label{sec:level3}Numerical Results}

\subsection{\label{sec:level4}The stationary state}
In the numerical calculation of the Lyapunov instability of a many-body system \cite{poschhoover}, there are some important time scales to be kept in mind: the first one is
the time for a many-body system to relax to a stationary state, which guarantees that quantities measured afterwards are not for a transient state;
the second is the time for the set of Lyapunov vectors to relax to their right orientations since offset vectors are
usually selected randomly at the beginning; the third is the time used to count LEs and LVs, which should be long enough to ensure that
 the trajectory wanders all over the
attractor. For a many-body system like the one studied here, these time scales can be extremely long due to the large number
of degrees of freedom involved \cite{pikovsky1,dellago}. 
\begin{figure}
\includegraphics*[scale=0.3]{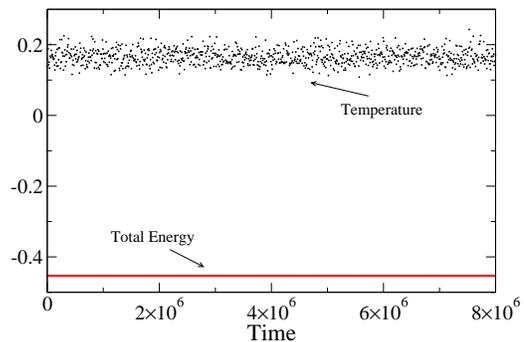}
\caption{\label{fig:stat} Time evolution of temperature $T \equiv \langle mv^2 \rangle$ and total energy.
 The nearly constant state variables show that the system has reached already a stationary state. 
 The parameter setting used here is: $N=100$, $L=1000$ and $T=0.2$.}
\end{figure}
\begin{figure}
\includegraphics*[scale=0.3]{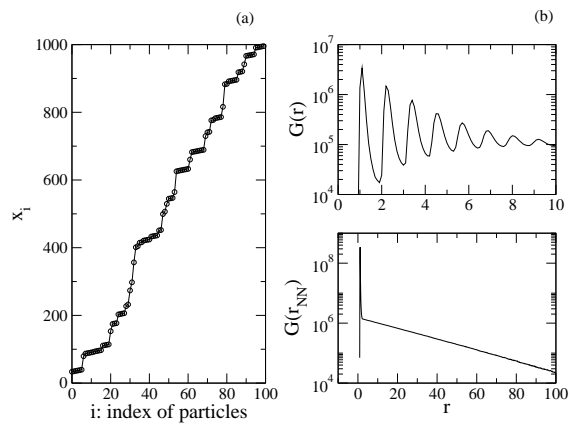}
\caption{\label{fig:gr} a) Particle positions $x_i$ vs. index of particles $i$ and b) pair distribution function $G(r)$ for the space among all particles
(upper panel) and nearest neighbors (lower panel) for the stationary state shown in
Fig.~\ref{fig:stat} (see \cite{hy} for the definition of $G(r)$). The sharp peaks in $G(r)$ imply that the state is a broken-chain one. }
\end{figure}

The time evolution of state variables like temperature $T$ and total energy for a case with parameter setting $N=100$, $L=1000$ and $T=0.2$ is shown in 
Fig.~\ref{fig:stat}. In the beginning of our molecular dynamics simulation, particles are placed randomly in the interval $[0,L]$ . Their velocities are chosen 
randomly from a
Boltzmann distribution. In order to equilibrate the system, it is coupled to a stochastic heat bath
with given temperature $T$, i.e., every $500$ steps the velocities of the particles are replaced with velocities that were drawn from a Boltzmann distribution 
corresponding to that temperature.
 This was done for a time period of length $t_{eq}$, which is longer than the relaxation time of the system at this temperature. After the equilibration
procedure, the system is allowed to evolve with constant total energy, i.e., without the heat bath, for a time period of the same length as $t_{eq}$, 
in order to be sure
that the system is already in a stationary state at given temperature $T$. In Fig.~\ref{fig:stat},
the period with thermal bath is omitted and only the part of evolution with constant total energy is shown. 
The nearly constant value of temperature means that the system has already
reached a stationary state and one can start the calculation of the Lyapunov instability of the system. The pair distribution function $G(r)$ shown in
Fig.~\ref{fig:gr} tells us that
 the stationary state for $T=0.2$ is a broken-chain state with short range order. 
 This is generic for a 1d Lennard-Jones system with not too high density \cite{stillinger}. 

\subsection{Benettin method using Gram-Schmidt orthogonalization}
The standard method invented by Benettin and Shimada \cite{benettin,shimada} is the most efficient one to calculate the 
Lyapunov exponents and Lyapunov vectors of large systems. Here 
$N\times N$ linearized equations for offset vectors in the tangent space were integrated simultaneously with the set of $N$
nonlinear equations for the reference trajectory.
Offset vectors were periodically re-orthonormalized using the Gram-Schmidt algorithm. 
The resulting rescaling factors measure the expansion or contraction rate of offset vectors in certain directions. Averaging
the logarithm of them for a period $\tau$ gets what are called {\it finite-time Lyapunov exponents $\lambda_{\tau}$}. 
The limit $\lambda \equiv \lambda_{\tau \to +\infty}$ is what is usually called {\it Lyapunov exponent}. 
The value of finite-time LEs $\lambda_{\tau}$
depends on the trajectory segment where it is calculated and usually it fluctuates as the segment moves along the trajectory. However
$\lambda_{+\infty}$ is time independent and unique for an ergodic system. 
In this sense, $\lambda_{+\infty}$ is a global quantity in characterizing the system attractor while the finite-time LEs are local
quantities which contain more detailed information about the dynamics. 
The offset vectors just after re-orthonormalization
are called {\it Lyapunov vectors}. It is a local quantity in characterizing the system attractor similar to the finite-time LEs.  

Another point to be noted is that Lyapunov vectors obtained using Benettin's method are always mutually orthogonal 
while the local unstable and stable directions are in
general not orthogonal. In this sense, these are two different sets of vectors. They are also different from the one in the multiplicative
ergodic theorem \cite{ergo}. Lyapunov vectors obtained in the standard way can at least represent the most unstable direction in a certain
subspace and they already contain a lot of important information about the dynamical instability in tangent space.
We will rely on them to continue our study in this paper. 

\subsection{Lyapunov exponents with wild fluctuations}
\begin{figure}
\includegraphics*[scale=0.3]{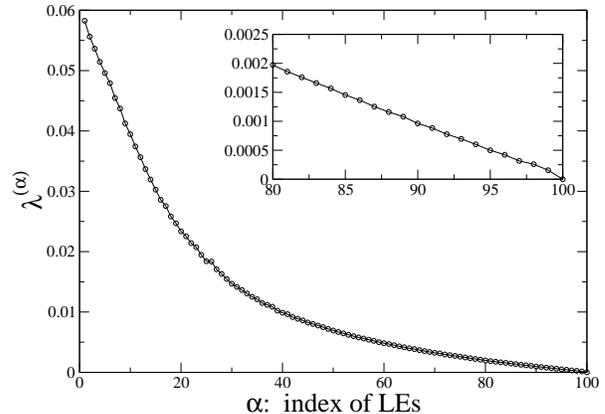}
\caption{\label{fig:lep-0.166} Lyapunov spectrum for the state shown in Fig.~\ref{fig:gr}. Enlargement of the part in the regime $\lambda^{(\alpha)} \approx 0$
is shown in the inset. It is the result of an average over $58$ samples with different initial conditions and for each sample the averaging period is $4 \times 10^6 h$.
 Here no step-wise structure exists in contrast to the case of hard-core system. This is the typical result for our soft potential system. 
}
\end{figure}
The Lyapunov spectrum for the case with $N=100$, $L=1000$ and $T=0.2$ is shown in Fig.~\ref{fig:lep-0.166}. Here only half of the spectrum is shown since
all LEs of Hamiltonian systems come in pairs according to the conjugate-pairing rule \cite{djevans,dettmann}.
From the enlargement shown in the inset of Fig.~\ref{fig:lep-0.166} for the part near $\lambda^{(\alpha)} \approx 0$, one can not see any step-wise structure 
in the Lyapunov spectrum in contrast to the case of hard-core
systems \cite{posch2002}. This is the typical result obtained for our soft potential system.
\begin{figure}
\includegraphics*[scale=0.3]{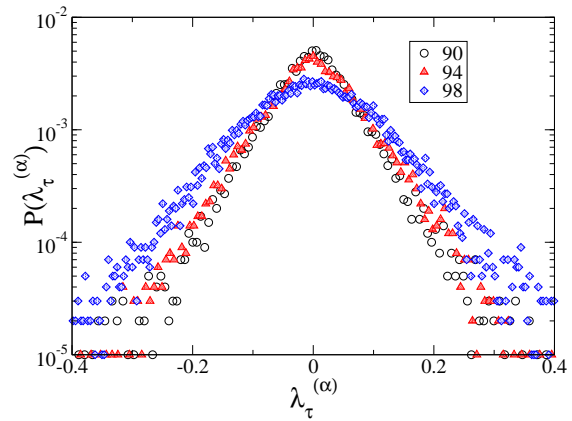}
\caption{\label{fig:lep-flu} Distribution of the finite time Lyapunov exponent $\lambda^{(\alpha)}_{\tau}$ where $\tau$ is equal to
the period of re-orthonormalization and the index of LEs
$\alpha$ is equal to $90$, $94$ and $98$ respectively. The strong fluctuations of $\lambda^{(\alpha)}_{\tau}$ are one of the possible reasons for the disappearance
 of the step-wise structures in the Lyapunov spectrum. The parameter setting used here is: $N=100$, $L=1000$ and $T=0.2$.}
\end{figure}

The fluctuations in local instabilities is demonstrated by the distribution of finite-time LEs. In Fig.~\ref{fig:lep-flu} such distributions for some LEs in the regime $\lambda
\approx 0$ are presented. Fluctuations of the finite time Lyapunov exponents are quite large compared with the difference between their mean values, i.e., 
$\sigma(\lambda^{(\alpha)}_{\tau})\equiv \sqrt{ \langle {\lambda^{(\alpha)}_{\tau} }^2 \rangle-{\langle \lambda^{(\alpha)}_{\tau}  \rangle}^2 }\gg |\lambda^{(\alpha)}-\lambda^{(\alpha+1)}|$. Here $\langle \cdots \rangle $ means time average.
The strong fluctuations in local instabilities is one of the possible reasons for the disappearance of the step-wise structures in the Lyapunov spectra. 
 It could also cause the mixing
of nearby Lyapunov vectors. The mixing may be at the origin of the intermittency observed in the time evolution of the spatial Fourier transform of LVs (see Sect.
\ref{sec:lv}).

\begin{figure}
\includegraphics*[scale=0.3]{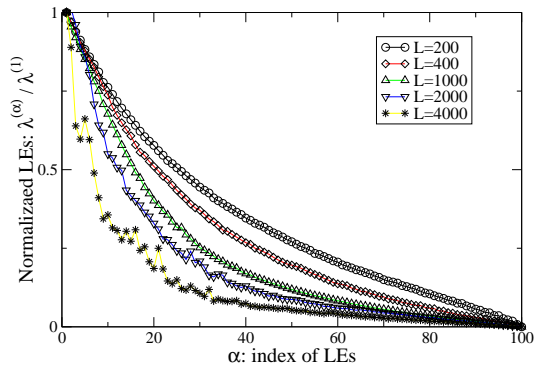}
\caption{\label{fig:le-rou} Normalized Lyapunov exponents $\lambda^{(\alpha)}/\lambda^{(1)}$ with $L=200$, $400$, $1000$ $2000$, and 
$4000$ respectively. The Lyapunov spectrum becomes more and more bended as the particle density $\rho=N/L$ is decreased. This implies the separation of two time scales.
Here $N=100$ and $T=0.2$.}
\end{figure}
\subsection{Bending of Lyapunov spectrum with decreasing particle density}
We studied also the influence of the particle density on the Lyapunov spectrum by increasing the length of the system with the particle
number $N$ kept fixed. As can be seen in Fig.\ref{fig:le-rou}, the Lyapunov spectrum becomes more and more bended with increasing $L$. For the
case of $L=4000$, the spectrum can already be unambiguously divided into two regimes: In the upper regime, Lyapunov exponents decrease more quickly with increasing index than
in the lower regime. This bending of the Lyapunov spectrum was related to the separation of two time scales in dilute particle
systems \cite{morriss2003}. We propose that one is the time of local collision events, and the other is due to the collective motion of the particles. 
For a system with high density, the
collisions are so frequent that there are strong correlation between consecutive collisions and
one can no longer separate them from each other. The collisions themselves contribute to the collective motion
of the system. Therefore no time scale separation happens here and the LEs decrease gradually. 

\subsection{\label{sec:lv}Spatial structure of LVs with $\lambda^{(\alpha)} \approx 0$}
\subsubsection{Coordinate fluctuation density (CFD)}
Another quantity used to characterize the local instability of trajectories are Lyapunov vectors $\delta \Gamma^{(\alpha)}$, which represent expanding or contracting 
directions in tangent space. 
In the study of hard-core
systems, Posch et al. found that coordinate part of the Lyapunov vectors corresponding to $\lambda \approx 0$ are of regular wave-like
character \cite{posch-hirschl,PoschForster,posch2002}. They are referred to as {\it hydrodynamic Lyapunov modes}.
We are searching here for the counterpart of these modes in our soft-potential system. 

\begin{figure}
\includegraphics*[scale=0.3]{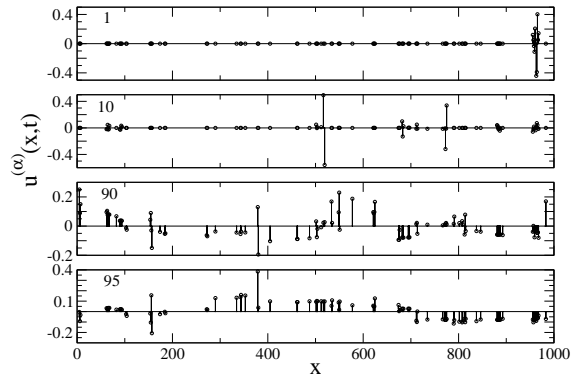}
\caption{\label{fig:lv} $u^{(\alpha)}(x,t)$ for LVs with index $\alpha=1$, $10$, $90$, and $95$ respectively. Notice that the former two are more localized while the latter two
are more distributed. The parameter setting used here is: $N=100$, $L=1000$ and $T=0.2$.}
\end{figure}
Remember that each of the LVs is consist of two parts: the displacement $\delta x_i$ in coordinate space and $\delta v_i$ in momentum space.
In past studies of hydrodynamic Lyapunov modes in hard-core systems, only the coordinate part $\delta x_i$ was considered. 
This is due to an interesting feature of LVs found in Ref.\cite{france} which says that 
the angles between the coordinate part and the momentum part are always small, i.e, the two vectors are nearly parallel. Therefore, it is already sufficient 
to use only $\delta x_i$ for studying $\delta \Gamma$. For our soft potential system, we find that the angles between the coordinate part and the momentum part
are no longer as small as in the hard-core systems. However, we will still
follow the tradition to study the coordinate part of LV first and come to the momentum part afterwards. 
\begin{figure}
\includegraphics*[scale=0.7]{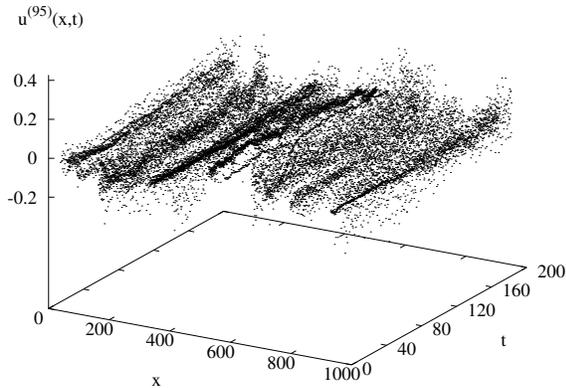}
\caption{\label{fig:lv-time} Time evolution of $u^{(95)}(x,t)$. No clear wave structure can be detected. The parameter setting used here is: $N=100$, $L=1000$ and $T=0.2$.}
\end{figure}

Analogous to the definition of microscopic densities in hydrodynamics \cite{hy},
we define a quantity called {\it coordinate fluctuation density}
\begin{equation}
u^{(\alpha)}(x,t)=\sum_{i=1}^N \delta x^{(\alpha)}_i \cdot \delta(x-x_i).
\end{equation}
%u^{(\alpha)}(x,t)=\frac{1}{\sqrt N}\sum_{i=1}^N \delta x^{(\alpha)}_i \cdot \delta(x-x_i).
%It is actually the coordinate part of LVs as function of the particle coordinate, the 
%quantity studied in the previous works \cite{posch2002,eckmann,france,wijin,Taniguchi,morriss}. 

Profiles of $u^{(\alpha)}(x,t)$ for some typical LVs of the Lennard-Jones
system are presented in Fig.~\ref{fig:lv}. It can be seen that $u^{(\alpha)}(x,t)$ for LVs corresponding to the largest Lyapunov exponents are highly localized,
 for example $u^{(1)}(x,t)$ and $u^{(10)}(x,t)$, while
those for $LV_{90}$ and $LV_{95}$ are more distributed. 
The study on the localization of $LV_1$ is of long history \cite{Pomeau,pikovsky3,pikovsky2,egolf,LVlocalizedMPS,PoschForster,posch2002,morriss2003} and
 it was related to defect events in simulations of Benard convection \cite{egolf}.
We leave the discussion on this point to Sec.\ref{sec:local}.
The temporal evolution of $u^{(95)}(x,t)$ is shown in Fig.~\ref{fig:lv-time} in order to make the possibly existing
wave-like structure more evident.
A wave structure however, cannot unambiguously be detected here with the naked eye.

\subsubsection{Spatial power spectrum of CFD and intermittency in its time evolution}
Now we turn to the spatial Fourier transform of $u^{(\alpha)}(x,t)$, which reads
\begin{equation}
 \begin{split}
\tilde{u}_k^{(\alpha)}(t) & =\int u^{(\alpha)}(x,t) \exp(-ikx)dx\\
                          & =\sum_{j=1}^N \delta x^{(\alpha)}_j \cdot \exp [ -ik \cdot x_j(t)] 
\end{split}
\end{equation}

In previous studies, in order to make the wave structure more obvious, certain smoothing procedures in time or space were applied
 to the Lyapunov vectors. For a 1d hard-core system with only a few particles, this procedure has been shown to be quite useful in identifying the
 existence of hydrodynamic Lyapunov modes \cite{morriss}. The success of this strategy relies on the fact that some of the Hydrodynamic Lyapunov modes (transverse modes) 
 of hard-core systems
 are stationary \cite{posch-hirschl}. Therefore, time averaging can indeed suppress the noise component and make the long wave-length modes more significant. 
 For a soft potential system, all the Lyapunov vectors are not stationary due to the random mixing among them. Especially, for our one dimensional system
 studied here, no transverse modes but only longitudinal Lyapunov modes exist. In consequence of this, 
  the smoothing procedure is no longer very helpful for detecting the hidden regular modes and can even damages them \cite{hoover}. 
Here we apply the spatial Fourier transformation to the instantaneous quantity $u^{(\alpha)}(x,t)$ instead. The
algorithm offered especially for unevenly distributed data is very suitable for our case \cite{num}.
 Furthermore, we take the long time average (and ensemble average) of the spatial Fourier spectrum 
 \begin{equation}
 s_{uu}^{(\alpha)}(k,t)\equiv |\tilde{u}_k^{(\alpha)}(t)|^2
 \end{equation}
 since it is expected that in $S_{uu}^{(\alpha)}(k)\equiv \langle s_{uu}^{(\alpha)}(k,t)\rangle$ the contribution of stochastic fluctuations will be averaged out while 
the information about the collective modes will be left and accumulated. The following
results show that this technique is quite successful in detecting the hidden collective modes. 
\begin{figure}
\includegraphics*[scale=0.3]{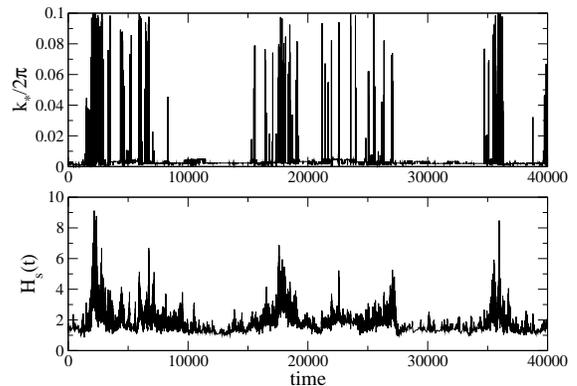}
\caption{\label{fig:k-en-time} Intermittent behavior of the peak wave-number $k_*$ and spectral entropy $H_s(t)$ for the spatial Fourier spectrum of $u^{(95)}(x,t)$.
 The parameter setting used here is: $N=100$, $L=1000$ and $T=0.2$.}
\end{figure}
\begin{figure}
\includegraphics*[scale=0.3]{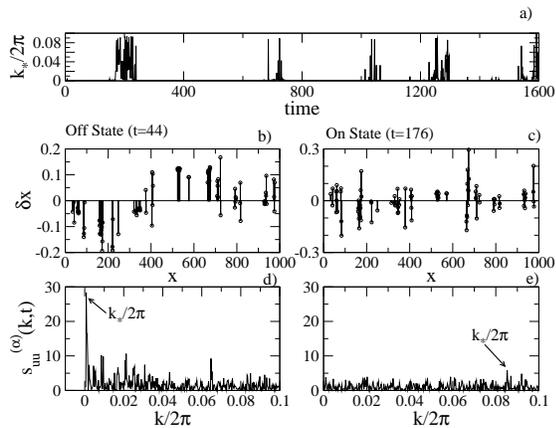}
\caption{\label{fig:lv-on-off} a) Variation of the peak wave number $k_*$ with time. b),c) Two typical snapshots of $LV_{95}$,
 {\it off} and {\it on} state at $t=44$ and $176$ respectively.
 d),e) their spatial
Fourier transform. The spectrum for the off state has a sharp peak at small $k_*$ while that for the on state has no dominant peak.
The parameter setting used here is: $N=100$, $L=1000$ and $T=0.2$.}
\end{figure}

The time evolution of the spatial Fourier spectrum $ s_{uu}^{(95)}(k,t)$ for Lyapunov vector No. $95$ is investigated in Fig.~\ref{fig:k-en-time} as an example.
Two quantities are recorded with time going on.
One is the peak wave-number $k_*$, which marks the position of the highest peak in the spectrum
$s_{uu}^{(\alpha)}(k,t)$ (see Fig. \ref{fig:lv-on-off}).
The other is the spectral entropy $H_s(t)$ \cite{entropy}, which measures the distribution property of the spectrum $s_{uu}^{(\alpha)}(k,t)$. 
It is defined as:
\begin{equation}
\label{eq_h}
H_s(t)=-\sum_{k_i}s_{uu}^{(\alpha)}(k_i,t)\ln s_{uu}^{(\alpha)}(k_i,t)
\end{equation}
A smaller value of $H_s(t)$ means that the spectrum $s_{uu}^{(\alpha)}(k,t)$ is highly
concentrated on a few values of $k$, i.e., these components dominate the behavior of the LV. Both of these
quantities behave intermittently as shown in Fig.~\ref{fig:k-en-time}. Large intervals of nearly constant low values ({\it off state}) are interrupted by short period of 
bursts ({\it on state}) 
where they experience large values. Details of typical {\it on} and {\it off} states are shown in Fig.~\ref{fig:lv-on-off}. 
It can be seen that the off state is dominated by the 
low wave-number components (see the sharp peak at low wave-number $k_*$) while the on state is more noisy and there are no significant dominant components. 
This intermittency in the time evolution of the spatial Fourier spectrum of LVs is a typical feature of soft potential systems.
 It is conjectured that this is a consequence of the mixing of nearby LVs caused by
the wild fluctuations of local instabilities. One time scale can be extracted from the intermittency, i.e., the mean duration $\tau_{off}$ of the {\it off} state. 
We conjectured that this time scale is related to the life time of hydrodynamics Lyapunov modes.
If the {\it off} state was viewed as a pure mode, the time $\tau_{off}$ is just the average life-time of such modes. Due to the mutual interaction among modes,
 the hydrodynamics Lyapunov modes in the soft potential systems are only of finite life-time. Further numerical work is needed in this direction to get more direct
 evidence.  

\begin{figure}
\includegraphics*[scale=0.3]{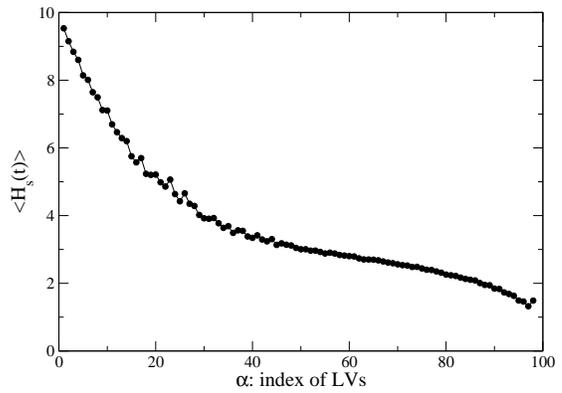}
\caption{\label{fig:lv-index-avg} Time-averaged spectral entropy $\langle H_s(t) \rangle$ vs. index of LVs. The gradual decrease of $\langle H_s(t) \rangle$ 
from $\alpha \approx 0$ means that LVs
corresponding to smaller positive LEs are more localized in Fourier space, i.e. they have more wave-like character those that corresponding to larger LEs. 
The parameter setting used here is: $N=100$, $L=1000$ and $T=0.2$.}
\end{figure}
In Fig.\ref{fig:lv-index-avg}, the time averaged spectral entropy $\langle H_s(t) \rangle$ is plotted against the index of the LVs.
 It increases gradually as the index decreases from $N-2$. This means that LVs
corresponding to smaller positive LEs are more localized in Fourier space, i.e., they have more wave-like character than those corresponding to larger LEs.
\subsubsection{Dispersion relation of hydrodynamical Lyapunov modes}
\begin{figure}
\includegraphics*[scale=0.45]{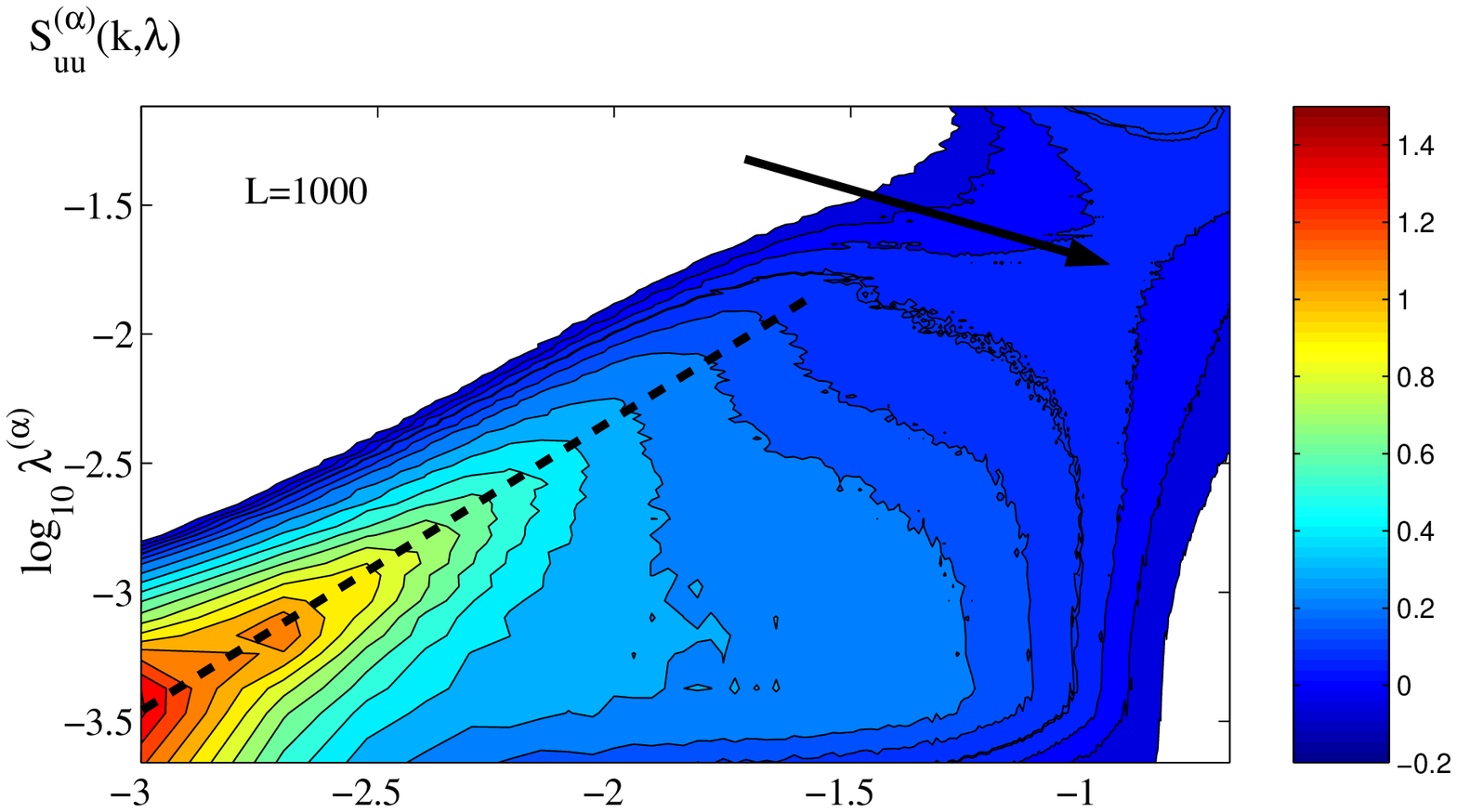}
\includegraphics*[scale=0.45]{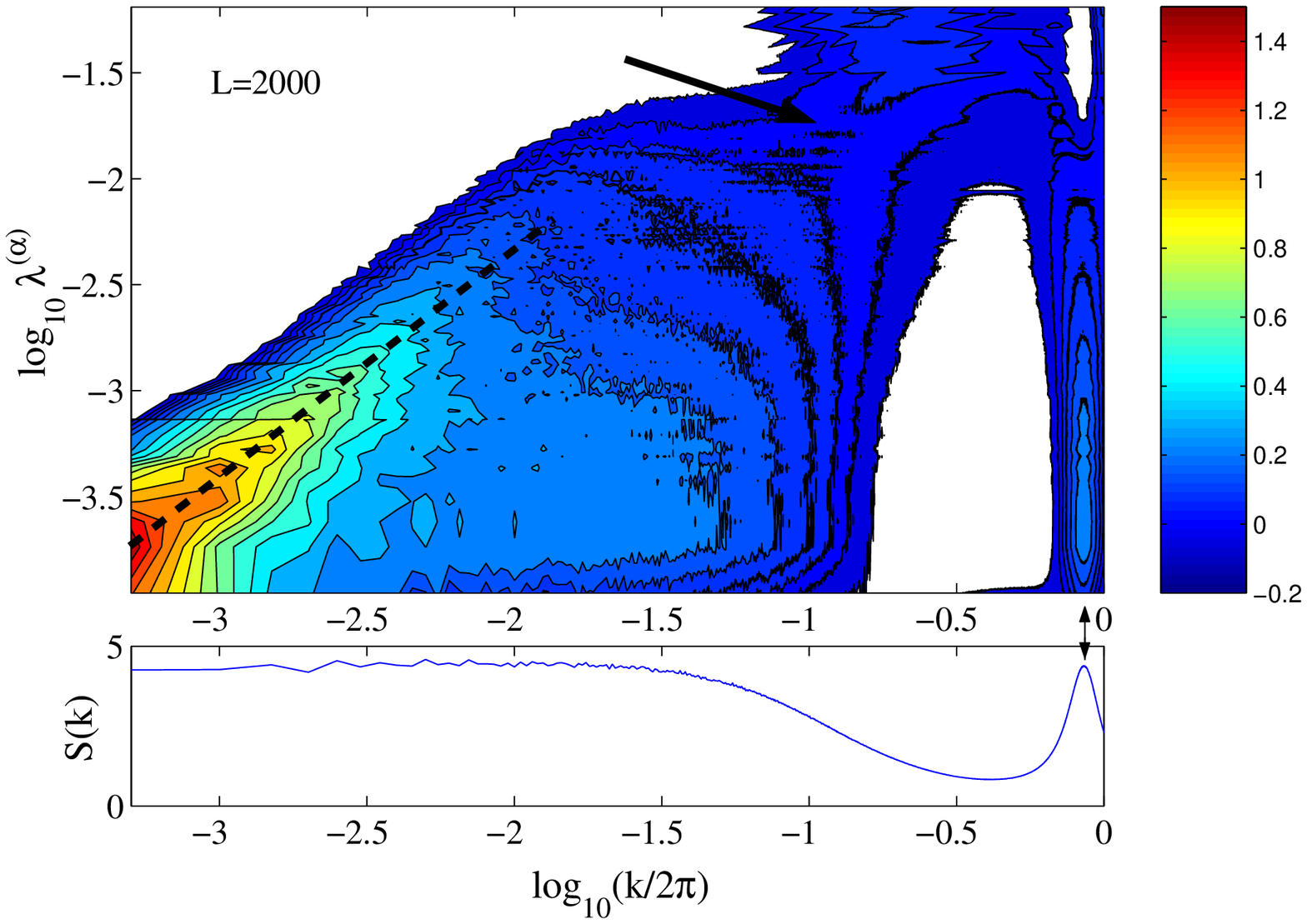}
\caption{\label{fig:uk-index-avg} (Color online) Contour plot of the spectrum $S_{uu}^{(\alpha)}(k)$ for $L=1000$ and $2000$. 
There is a sharp peak at $k \approx 0$ and $\lambda \approx 0$.
 To guide eyes, a dashed line is plotted to show how the peak wave-number $k_{max}$
changes with the variation of $\lambda$. The sudden jump in $k_{max}$ is marked with an arrow. In total $58$ samples for the case of $L=1000$ 
($10$ for $L=2000$) are used averaging each for a period of $4 \times 10^6 h$. Here $T=0.2$ and $N=100$.}
\end{figure}

\begin{figure}
\includegraphics*[scale=0.3]{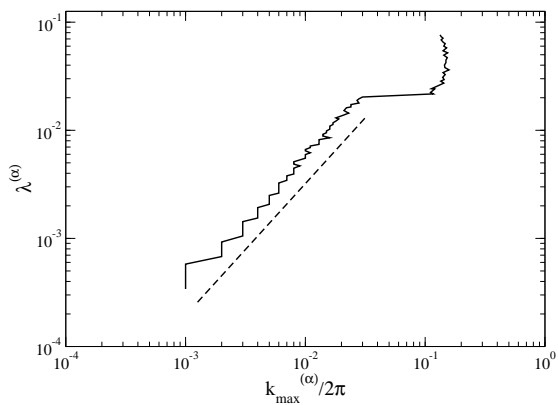}
\caption{\label{fig:lambda-k-only-1k}The Lyapunov exponent $\lambda^{(\alpha)}$ is plotted as function of the wave-number $k_{max}$ of the highest peak 
in the time averaged spatial Fourier spectrum of LVs. The dashed line is of the form
$ \lambda^{(\alpha)} \sim k_{max} ^{1.2}$. The parameter setting used here is: $N=100$, $L=1000$ and $T=0.2$.}
\end{figure}

Now we consider the time averaged spatial Fourier spectrum $S_{uu}^{(\alpha)}(k)$ of LVs. Two cases with $L=1000$ and $2000$ are shown in Fig.\ref{fig:uk-index-avg}. 
It is not hard to recognize the sharp peak 
at $\lambda \approx 0$ in the contour plot of the spectrum. In increasing the Lyapunov exponents, the peak shifts to the larger wave number side. A dashed line
is plotted to guide eyes how the wave number of the peak $k_{max}$ changes with $\lambda^{(\alpha)}$. To further demonstrate this point,
in Fig.\ref{fig:lambda-k-only-1k}, the value of the Lyapunov exponent $\lambda^{(\alpha)}$ is plotted versus $k_{max}$ of corresponding LVs. We call this the 
{\it dispersion relation} of 
the hydrodynamical Lyapunov modes. The numerical fitting of the data shows that for $\lambda \approx 0$, $\lambda^{(\alpha)} \sim k_{max}^ {\gamma}$ with
 the exponent $\gamma \approx 1.2$. However, a linear dispersion with quadratic corrections can not be excluded.
 
 In order to show that the peak in $S_{uu}^{(\alpha)}(k)$ is not a result of the highly regular packing of particles
in the broken-chain state, the static structure function \cite{hy}
\begin{equation} 
S(k) \equiv \int G(r) \exp(-ikr)dr 
\end{equation}
for the case $L=2000$ is plotted in the same figure as $S_{uu}^{(\alpha)}(k)$, where $G(r)$ is the pair correlation function shown in Fig. \ref{fig:gr}.
It can be seen that $S(k)$ is nearly constant in the regime $k\approx 0$, the place where a sharp peak was observed in $S_{uu}^{(\alpha)}(k)$. The regular packing of
particles causes the formation of a peak at $k/2\pi \approx 0.9$ in $S(k)$. This corresponds to a tiny peak at the same $k$-value in $ S_{uu}^{(\alpha)}(k)$ for those LVs
with $\lambda \approx 0$. These facts show clearly that the collective modes observed in LVs are not due to the regular packing of particles. 

\begin{figure}
\includegraphics*[scale=0.3]{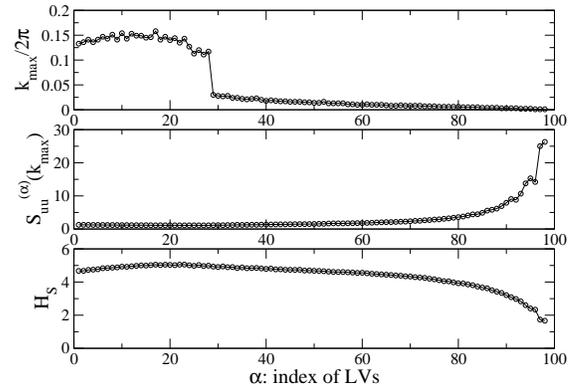}
\caption{\label{fig:lv-avg-T=0.2} (Upper panel) The wave-number $k_{max}$ of the highest peak 
in the time averaged spatial Fourier spectrum of LVs. (Middle panel) The height $S_{uu}^{(\alpha)}(k_{max})$ of the highest peak in the time
averaged spectrum. (Lower panel) The spectral entropy $H_{S}$ for the averaged spectrum $S_{uu}^{(\alpha)}(k)$.
The sudden jump in $k_{max}$ implies the separation of time scales.
The parameter setting used here is: $N=100$, $L=1000$ and $T=0.2$.}
\end{figure}
\begin{figure}
\includegraphics*[scale=0.3]{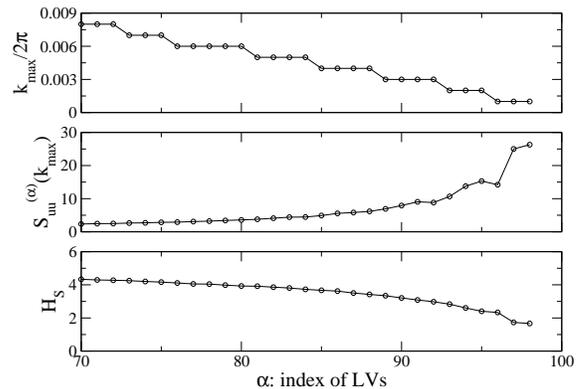}
\caption{\label{fig:lv-avg-T=0.2-b} Enlargement of Fig.\ref{fig:lv-avg-T=0.2} for the part in the regime $\alpha \simeq N$.}
\end{figure}
To further demonstrate the properties of these modes, in Fig.\ref{fig:lv-avg-T=0.2}, $k_{max}$ is plotted versus index of LVs. 
As can be seen, for LVs with
$\lambda \approx 0$, i.e. with $\alpha \approx N$, the value of $k_{max}$ is quite small (see the enlargement in Fig. \ref{fig:lv-avg-T=0.2-b}). For example,
for $\alpha=96$, $97$, and $98$, $k_{max}=2\pi/L$, which is the smallest wave number allowed by the periodic boundary conditions used. Another point to be noticed is the 
step structure in plotting $k_{max}$ as function of $\alpha$. This is similar to the degeneration of wave-numbers found in the hard-core case,
 although the steps here are not so regular. 
In the middle panel of Fig.\ref{fig:lv-avg-T=0.2}, the height $S_{uu}^{(\alpha)}(k_{max})$ of the highest peak in the time averaged spatial spectrum 
is plotted with index of LVs. Except LV No.99 and No.100 for the
conserved quantities, $S_{uu}^{(\alpha)}(k_{max})$ decreases gradually in decreasing the index from $N-2$. Similar to the definition of the spectral entropy for the instantaneous
spectrum $s_{uu}^{(\alpha)}(k,t)$ in Eq.\ref{eq_h}, one can also define a spectral entropy $H_{S}$ for the averaged spectrum $S_{uu}^{(\alpha)}(k)$. The spectral entropy $H_{S}$
for $S_{uu}^{(\alpha)}(k)$ is presented in the lower panel of Fig.\ref{fig:lv-avg-T=0.2}. It possesses a minimum at $\alpha=98$ where the Lyapunov
exponent is the smallest positive one. According to the definition of the spectral entropy, the minimum means that the spectrum of LV No.98 is
most significantly dominated by a few components.
 
All of our results shown above give strong evidence
 that the Lyapunov vectors corresponding to the smallest positive LEs in our 1d Lennard-Jones system are highly dominated by a few components with small wave numbers, i.e, 
they are similar to the Hydrodynamic Lyapunov modes found in hard-core systems. The wave-like character becomes weaker and weaker as the value of LE is increased gradually 
from zero.
 
\begin{figure}
\includegraphics*[scale=0.3]{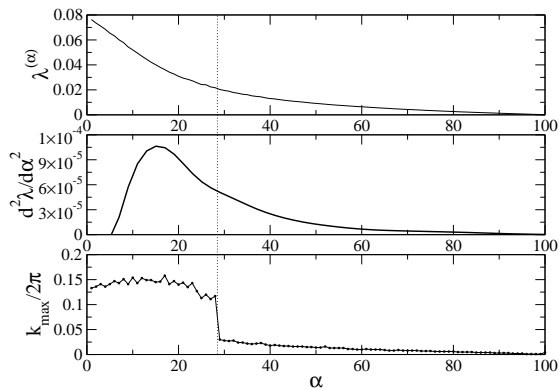}
\caption{\label{fig:k-index-lamb} The Lyapunov spectrum $\lambda^{(\alpha)}$ (upper panel),
$d^2 \lambda /d\alpha^2$ (middle panel) and $k_{max}$ vs. the index $\alpha$. The parameter setting used here is: $N=100$, $L=1000$ and $T=0.2$.}
\end{figure}
\subsubsection{Separation of time scales}
Another interesting point in Fig.\ref{fig:lv-avg-T=0.2} is the
sudden jump in $k_{max}$ at $\alpha \approx 26$ which divides the whole set of LVs into two groups. 
It is believed that this sudden jump is related with the bending of the LE spectrum and the separation of time
scales in a dilute system. As shown in 
Fig. \ref{fig:k-index-lamb}, the sudden jump is in the regime where the LEs spectrum is most strongly bended, although it is not at the exact place where
$d^2 \lambda^{(\alpha)} / d \alpha^2$ experiences the maximal value. Further work is needed to reveal the underlying
connection between these phenomena.        

\begin{figure}
\includegraphics*[scale=0.3]{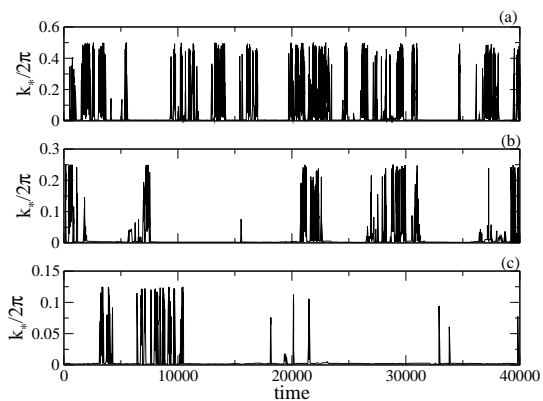}
\caption{\label{fig:k1st-density} Same as top panel of Fig. \ref{fig:k-en-time}, but with different density (a) $\rho=1/2$, (b) $1/4$ and (c) $1/8$ respectively.
 Here $T=0.2$ and the particle number $N=100$.}
\end{figure}
\begin{figure}
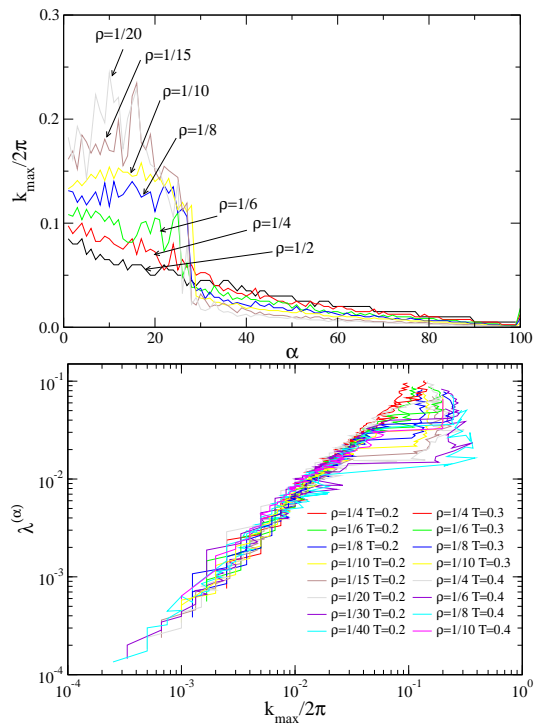

\includegraphics*[scale=0.3]{k-alpha-density-b.eps}
\includegraphics*[scale=0.3]{lambda-k-b.eps}
\caption{\label{fig:lv-avg-L}(Color online) (upper panel) The wave-number $k_{max}$ of the highest peak in the time-averaged spatial Fourier spectrum of LVs as function
of LV index $\alpha$. (lower panel) the Lyapunov exponent $\lambda^{(\alpha)}$ as function of $k_{max}$.
 Note that in the lower panel, all data from simulations with
different densities and temperatures collapse to a single curve. Fitting of the low wave-number part to a power-law function 
$ \lambda^{\alpha} \sim k_{max}  ^\gamma$ gives $\gamma \approx 1.2 \pm 0.1$. Here $N=100$ and $T=0.2$.}
\end{figure}
\subsubsection{Influence of density and temperature}
To study how the change in density influences the behavior of LVs, we increase the length $L$ of the system from $200$ to $4000$ with the particle number
$N$ kept fixed at $100$. From the time evolution of $k_*$ shown in Fig.\ref{fig:k1st-density}, one sees that, in increasing the density $\rho=N/L$, the
occurrence of the $\it on$-state becomes more frequent, i.e., the domination of low wave-number components is much weaker.
 The spatial Fourier spectrum for LVs with LEs in the regime $\lambda^{(\alpha)} \simeq 0$, however, are always dominated by certain low wave-number components
irrespective of the density (see Fig.\ref{fig:lv-avg-L}). 

An important point is the collapse of data of dispersion relations from simulations with various densities and temperatures to a single curve (see
lower panel of Fig.\ref{fig:lv-avg-L}). This means, for hydrodynamic Lyapunov modes in our system,
there is a universal dispersion function $\lambda_{\alpha} (k)$ irrespective to the particle density and the system temperature. 
%This is one of the main results of this study.
 Fitting of the data to a power-law function $\lambda_{\alpha} \sim k_{max}  ^\gamma$ says the value of the exponent $\gamma$ is $1.2 \pm 0.1$ although a
linear dispersion with quadratic corrections cannot be excluded. 

Another feature of Fig.\ref{fig:lv-avg-L} (upper panel) is that the sudden jump in $k_{max}$ disappears as the system density $N/L$ is increased. 
This is consistent with our above discussion that the separation of time scales is only significant in dilute systems. 

\subsubsection{Dynamics of the momentum part}
Now we turn to investigations on the spatial Fourier spectrum of the momentum part of LVs. Unfortunately, all the spectra are
more or less homogeneously distributed on all wave-numbers. For all the cases checked, no wave-like structure can be identified as for the coordinate part.
 One may wander why no mode-like collective motion is observed in the momentum part. There are two possibilities, one is that
the momentum part does contain information similar to the coordinate part but it is too weak to be detected here due to the strong noise.  
The other is that there is no similarity between the
two parts at all and regular long wave-length modes exist only in the coordinate part. Further work is needed to clarify which one is correct.

\begin{figure}
\includegraphics*[scale=0.3]{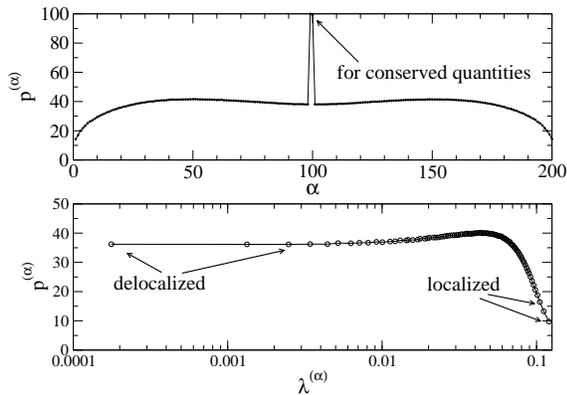}
\caption{\label{fig:p-i} Time-averaged participation number $p^{(\alpha)}$  vs. index of
LVs (upper panel) and $\lambda^{(\alpha)}$ (lower panel). The parameter setting used here is: $N=100$, $L=110$ and $T=1.6$.}
\end{figure}
\subsection{\label{sec:local}Localization properties of LVs}

To study the localization of LVs, we employ the participation number, which is defined as: 
\begin{equation}
p \equiv \bigg[ \sum_{i=1}^N (\delta x_i^2+\delta v_i^2)^2 \bigg]^{-1}
\end{equation}
for a Lyapunov vector $(\delta x^{(\alpha)}_i,\delta v^{(\alpha)}_i)$ \cite{note_pik}. This is a standard quantity used in the study of disorder-induced 
localization \cite{par}, which roughly measures the number of particles
which contribute to the Lyapunov vector. For the homogeneous Lyapunov vector $LV_{N}$ with $w^2\equiv \delta x_i^2+\delta v_i^2 =1/N$, which corresponds to one of the
zero-value LEs, $p$ gets its maximal value $N$. In decreasing the index of LVs, LEs become larger and larger. Accompanying to this,
the participation number $p^{(\alpha)}$ decreases as presented in Fig.~\ref{fig:p-i}, where the variation of the time-averaged value 
of $p^{(\alpha)}$ versus index $\alpha$ and $\lambda^{(\alpha)}$ are plotted. The decrease of $p^{(\alpha)}$ implies that LVs become more
and more localized with decreasing $\alpha$.
\begin{figure}
\includegraphics*[scale=0.3]{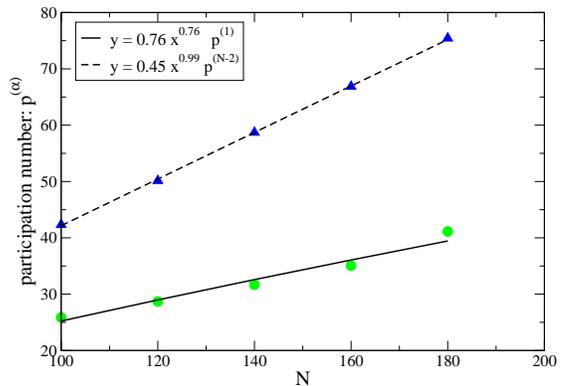}
\caption{\label{fig:p-L} Time-averaged participation number $p^{(1)}$ and $p^{(N-2)}$ vs. particle number $N$.
 The parameter setting used here is: $L/N=1.1$ and $T=1.6$.}
\end{figure}
\begin{figure}
\includegraphics*[scale=0.3]{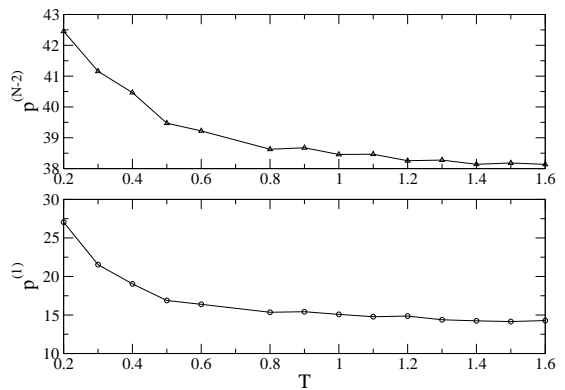}
\caption{\label{fig:p-t} a) Time-averaged participation number $p^{(1)}$ and $p^{(N-2)}$ vs. temperature $T$. The parameter setting used here is: $N=200$ and $L=220$.}
\end{figure}

One should note
that $p^{(N-2)}$ for $LV_{N-2}$ corresponding to the smallest positive Lyapunov exponent is significantly different from $p^{(N)}=N$ for $LV_{N}$ with $\lambda^{(N)}=0$. In
Fig.~\ref{fig:p-i}, $p^{(N-2)} \approx 40$ while $p^{(N)}=100$.
In the study of space-time chaos, a commonly used
measure for quantifying spatio-temporal disorder is the fractal dimension $D$. A spatio-temporal chaotic system is generally extensive, with the fractal dimension $D$ in proportion to
its volume $V$. According to this, a bounded intensive quantity named dimension correlation length is defined $\xi_{D}=\lim_{V\to \infty}(D/V)^{-1/d}$. 
 Based on the intuitive thought that
a spatio-temporal chaotic system is composed of many subsystems and that these subsystem are uncorrelated if they are far apart, it is
expected that $\xi_D$ is proportional to the two-point correlation length, which measures the spatial disorder of the system.
 A particular fractal dimension, 
the Lyapunov dimension $D_{\mathcal{L}}$ can be easily obtained using the Kaplan-Yorke formalism which relates the Lyapunov dimension to Lyapunov
exponents of the system \cite{kp}. 
 For a Hamiltonian system like the one studied here, Lyapunov exponents are paired, i.e. $\lambda^{(\alpha)}+\lambda^{(2dN-\alpha)}=0$ due to the sympletic structure of
 the system \cite{djevans,dettmann}. 
 According to the Kaplan-Yorke formalism, the
 Lyapunov dimension for a $d$-dimensional Hamiltonian spatio-temporal chaotic will be $D_{\mathcal{L}}=2dN$ in spite of the details of dynamics.
 Although $D_{\mathcal{L}}$ defined in such way is proportional to the volume of the system i.e., reflects the extensiveness of the system, its value is always a
 constant irrespective of the temperature change. However the two-point correlation length 
 does change with the temperature and it even becomes divergent as a phase transition is encountered. In this sense, the Lyapunov dimension $D_{\mathcal{L}}$ and
  the dimension correlation 
 length $\xi_{D}$ are only trivially defined here and not good quantities to characterize spatio-temporal Hamiltonian chaos. 
 Here we propose a new length scale based on the
 participation number of LVs:
 \begin{equation}
 \xi^{(\alpha)}_{p}\equiv \bigg ( \frac{p^{(\alpha)}}{N}\cdot V \bigg )^{1/d}
 \end{equation}
 where $d$ is the dimension of the physical space and $V$ is the system volume which is simply $L$ for our $d=1$ case here.
According to the fact we mentioned above, $\xi^{(N-2)}_p$ for the smallest positive Lyapunov exponent is a nontrivial value depending on the state of the system. 

Now we first check the extensiveness of $\xi^{(N-2)}_p$. 
 The fit of the numerical data in Fig.~\ref{fig:p-L} gives $p^{(N-2)} \sim N^{0.99}$. 
 This tells us $p^{(N-2)}$ (and consequently $\xi^{(N-2)}_p$) is proportional to the system size $N$, i.e., it is an extensive quantity.
 The dependence of $p^{(1)}$ for the largest Lyapunov exponent on $N$ is shown in the same plot. It is fitted with $p^{(1)} \sim
N^{0.76}$, i.e., the Lyapunov vector for the largest LE is highly localized in space \cite{pikovsky2,morriss2003,posch2002}.
 This is consistent with our above observation (see Fig.~\ref{fig:lv}).

Then we study the temperature dependence of this newly defined length scale. For a $200$-particle system, the temperature is increased from $0.4$ to $1.6$. Results of the
simulation are shown in Fig.~\ref{fig:p-t}, where the variation of $p^{(1)}$ with temperature $T$ is also presented.
 From the plot, one can see that $p^{(N-2)}$ (and consequently $\xi^{(N-2)}_p$) decreases gradually with the
increase in temperature. This agrees with the intuitive expectation that, increasing of the temperature makes the fluctuation in the system stronger and stronger and 
renders the system becoming more and more disordered. 

\section{Conclusion and Discussion}
In this paper, we presented numerical results about the Lyapunov instability of a Lennard-Jones system. Our simulations show that the step-wise structure 
found in the Lyapunov spectrum of hard-core systems disappears completely here. This is conjectured due to the strong fluctuations in the
finite-time LEs \cite{posch2002}. A new technique \cite{radons} based on a spatial spectral analysis is employed to reveal the hidden
long wave-length structure in LVs. A significantly sharp peak with low wave-number is found in the resultant spatial Fourier spectrum for LVs with $\lambda \simeq 0$.
This serves as strong evidence that hydrodynamic Lyapunov modes do exist in the soft-potential systems \cite{posch2004}.
Another important finding is that the dispersion relation for hydrodynamic Lyapunov modes, $\lambda^{(\alpha)}$ versus $k_{max}$, appears to be universal irrespective
of system temperature and particle density. 

In the study of two-dimensional and quasi-one-dimensional hard-core systems, two kinds of hydrodynamic Lyapunov modes are identified \cite{posch2002,morriss}.
One is referred to as transverse and the other is called
longitudinal. The former modes do not propagate while the latter can \cite{posch2002}. According to this classification, for the transverse Lyapunov modes,
taking the time average can be an useful way to identify the wave-like structure. In contrast to this, the detection of the longitudinal Lyapunov modes
is a relatively difficult task since due to its propagation time-averaging is no longer a suitable method to suppress the fluctuations \cite{morriss}. For the case of soft potential
system, strong fluctuations in local instabilities leads to the occasional mixing among Lyapunov vectors. This is partially reflected in the intermittent time evolution of
spatial Fourier spectrum of LVs. 
Therefore the hydrodynamic Lyapunov modes in a soft-potential system are more vague and more difficult to observe \cite{hoover}. 

In the study of the hard-core case, it is conjectured that 
degeneracies in the Lyapunov spectrum and in wave-numbers of hydrodynamic Lyapunov modes are determined by the intrinsic symmetries of the Hamiltonian and the boundary conditions.
There is no contradiction between this statement and the results reported here. The crucial point is the life-time of hydrodynamic Lyapunov modes. The
above statement is for the ideal
case of pure modes with infinite long life-time. For the Lennard-Jones system studied here, strong fluctuation in local instability leads to the mutual interaction and mixing
among modes, which renders the life-time of modes becoming finite. On the other hand, Lyapunov exponents are global quantities of the system, which are the result of a time-average 
along a
long trajectory wandering all over the phase space allowed. Due to the mixing among modes, they no longer correspond to certain pure modes but to a mixture of several modes.
Therefore the degeneration predicted on the basis of a symmetry analysis can not be seen here. 
Actually, fluctuations in local instabilities do exist
in all dynamical systems. However, for the hard-core systems, it is relatively weak in the regime of $\lambda \approx 0$ (see Fig.12 in \cite{posch2002} 
for the comparison of fluctuations in local instabilities for the two cases). There the mixing among modes is quite rare and weak. The life-times of the modes with $\lambda
\approx 0$  are so long that one can not feel it during the finite simulation time. Evidence for our argument above comes from the fact that as the fluctuations in local instabilities
of hard-core system become stronger and stronger in increasing LEs from $\lambda \approx 0$, the degeneration in LEs becomes worse and worse, i.e. the steps become
 steeper and steeper (see Fig.12 and Fig. 8 in Ref.\cite{posch2002}). In this work, Fourier spectral analysis has been shown to be quite successful in 
 detecting the hidden 
wave-like structures in LVs. A more general theoretical consideration of this method will be given elsewhere \cite{radons}. 

Until now, only the coordinate part of the LVs are used in the study of hydrodynamic Lyapunov modes. 
For the case of hard-core system, this is reasonable due to the interesting feature of LVs found in \cite{france} that 
the angles between the coordinate part and the momentum part are always small, i.e, the two vectors are nearly parallel. 
For our soft potential system, we find that the angles between the coordinate part and the momentum part
are no longer as small as in the hard-core systems. Even for the hard-core system, recent results show that the two vectors are not always parallel \cite{morriss2003}. 
Therefore it is necessary to reconsider the momentum part of LVs. Due to the specific feature of Hamiltonian systems, the momentum part of a LV with $\lambda>0$ is identical to
the coordinate part of a LV with $-\lambda$. This ensures that the study of only the coordinate part of LVs is sufficient. But now one need to deal with LVs corresponding to
both positive and negative LEs. In general, the behavior of LVs with negative LEs should be different from those with positive ones. For the case reported in the current paper, there
is no wave-like structure being observed in the former.

We studied also the influence of density and temperature changes on the features of LVs and LEs. 
One effect of decreasing the density is that the Lyapunov spectrum becomes more and more bended. The relation of the bending in the Lyapunov spectrum with the separation
of time scales was discussed recently in the work of Taniguchi and Morriss \cite{morriss2003}. It is obvious that,
 the collisions between particles become more and more
rare as the density is decreased. They turn to be highly localized events since they happen at only a few places at one moment.
Therefore, the time scale of local collision events, which is related to the largest LEs \cite{krylov}, is well separated from that for the collective
motion of the system, corresponding to the near-zero LEs. A further point is the sudden jump found in $k_{max}$ of the time-averaged spatial Fourier spectrum
of LVs. It divides the whole set of LVs into two groups. 
The place of this sudden jump is in the regime where the Lyapunov spectrum is strongly bended.
 Further work is needed to reveal the underlying connection between these phenomena.

\begin{acknowledgments}
We thank  H. A. Posch, W. Kob, A. S. Pikovsky, W. Just and A. Latz for fruitful discussions and W. Kob in addition for providing us with his MD simulation code, on which
our program is built. Financial support from SFB393 within the project "Long time behavior of large
dynamical systems" is gratefully acknowledged.
\end{acknowledgments}
%\bibliographystyle{unsrt}
%\newpage %Just because of unusual number of tables stacked at end
%\bibliography{lj1}% Produces the bibliography via BibTeX.

\begin{thebibliography}{50}
\bibitem{note_pik}
Here we follow Ref. \cite{pikovsky2} to construct a scalar variable $\delta x_i^2 + \delta v_i^2$ to study the localization of LVs.
\end{thebibliography}

\end{document}